\documentclass[
 amsmath,amssymb,
 twocolumn,
]{revtex4}

\usepackage{graphicx}

\begin{document}


\title{Delivering pulsed and phase stable light\\to atoms of an optical clock}

\author{Stephan Falke}
 \email{stephan.falke@ptb.de}
\author{Mattias Misera}%
\author{Uwe Sterr}%
\author{Christian Lisdat}%
\affiliation{Physikalisch-Technische Bundesanstalt, Bundesallee 100, 38116 Braunschweig, Germany\\fax: +49\,531\,592\,4305}


\begin{abstract}
In optical clocks, transitions of ions or neutral atoms are interrogated using pulsed ultra-narrow laser fields. Systematic phase chirps of the laser or changes of the optical path length during the measurement cause a shift of the frequency seen by the interrogated atoms. While the stabilization of cw-optical links is now a well established technique even on long distances, phase stable links for pulsed light pose additional challanges and have not been demonstrated so far. In addition to possible temperature or pressure drift of the laboratory, which may lead to a Doppler shift by steadily changing the optical path length, the pulsing of the clock laser light calls for short settling times of stabilization locks. Our optical path length stabilization uses retro-reflected light from a mirror that is fixed with respect to the interrogated atoms and synthetic signals during the dark time. Length changes and frequency chirps are compensated for by the switching AOM. For our strontium optical lattice clock we have ensured that the shift introduced by the fiber link including the pulsing acousto optic modulator is below $2\cdot 10^{-17}$.
\end{abstract}
\pacs{Valid PACS appear here}
\maketitle

\section{Introduction}
Time and frequency measurements have largely benefited from moving to higher frequencies in order to obtain lower uncertainties and better stability. The current definition of the second uses a Cs reference transition in the microwave regime \cite{cgp83}. It is expected that a re-definition of this best realized SI unit will bring the frequency up into the optical regime. Various references are under investigation to find candidates for the next time definition \cite{gil05}. In all these setups well prepared atoms or ions are interrogated by ultra-stable lasers with light pulses of a specific duration. The signal is used to lock the laser frequency to the center of the reference line, and the oscillations of the light field are compared with a frequency comb generated from a femtosecond laser \cite{ude02} to other rf or optical standards.

The past years have seen a steady progress \cite{boy07,lod10} in the field of optical clocks, which calls for optical links that are able to transfer the stability, both, between different optical clocks or between a clock laser and the reference atoms. The stability of the clock laser light needs to be maintained during its transfer from its stability providing reference cavity to the interrogated atoms. This task is complicated by the fact that the laser light for the interrogation of the atoms or ions needs to be switched to create interrogation pulses. Table~\ref{tab:drifts} lists a number of possible perturbations of the link and their related estimated shifts of the measured frequency. Any effect that causes the light to have one extra oscillation per second, for example by extending the beam path by one wavelength, leads to a frequency shift of 1~Hz. Perturbations as small as 3~nm/s therefore cause a shift on the order of 10$^{-17}$, which is a level that optical clocks are reaching. In addition, the generation of light pulses can also introduce a phase chirp \cite{deg05}.

The frequency of clocks can be realized not only with small systematic uncertainty but also with a high stability. Because of the Dick-effect, aliasing of fluctuations of the clock laser frequency considerably decreases the signal-to-noise ratio and thus degrades the stability of an optical clock \cite{dic87}. One can circumvent this problem by interrogating two ensembles of clock atoms with the same clock pulses. With optical lattice clocks instabilities as low as $1\cdot 10^{-17}$ in 1600 seconds have been observed with such a correlated interrogation scheme \cite{tak11}. Perturbations of the optical links to the two interrogated ensembles however will not be canceled, motivating the need of a stable optical link to achieve the ultimate performance of optical lattice clocks. With improved lasers, this will be also an important issue for the stability of a single clock without a correlated interrogation scheme.

\begin{table}
\begin{center}
\begin{tabular}{@{}lrrrr@{}}
\hline
\hline
effect %
& \hspace{1ex}change %
& \hspace{1ex}time %
& \hspace{1ex}length %
& \hspace{1ex}fract. shift\\
\hline
{\bf air pressure}%
&  %
&  %
&  %
&  \\
\hspace{0.5em} weather%
& 300 Pa %
& 1 h %
& 3 m%
& $2\cdot10^{-18}$ \\
\hspace{0.5em} laboratory door%
& 100 Pa %
& 3 s%
& 3 m%
& $9\cdot10^{-16}$ \\
{\bf temperature}%
& %
& %
& %
&  \\
\hspace{0.5em} air%
& 2~K %
& 1 h %
& 3 m %
& $5\cdot 10^{-18}$ \\
\hspace{0.5em} alumium%
& 2~K %
& 1 h %
& 1 m %
& $4\cdot 10^{-17}$ \\
\hspace{0.5em} expansion optical fiber%
& 2~K %
& 1 h %
& 20 m %
& $2\cdot 10^{-17}$ \\
\hspace{0.5em} refractive index fiber%
& 2~K %
& 1 h %
& 20 m %
& $4\cdot 10^{-16}$ \\
{\bf vibrations} see Fig.~\ref{fig:vibrations}%
& %
& %
& %
&  \\
\hspace{0.5em} mirror mount%
&\multicolumn{3}{l}{1~kHz, amplitude $\lambda$ }
& up to $10^{-16}$ \\
\hline
\hline
\end{tabular}
\end{center}
\caption{\label{tab:drifts}Order of magnitude estimations of various effects on the optical path length. Vibrations of mirrors may lead to significant effects if they are synchronous with the interrogation cycle as discussed in the text.}
\end{table}

In this paper we describe and investigate a stabilized optical link between an ultra-narrow clock laser and ultracold $^{87}$Sr atoms in an optical lattice including a switching acousto-optic modulator (AOM) that pulses the interrogation light. Typically fiber length stabilizations use light reflected from a reference point (like a fiber tip) that is not referenced to the atomic position. Often, the switching of the clock laser light is only provided in a short uncompensated path close to the atoms. 

In contrast to permanently stabilized fiber links in this work, the switching of the clock laser light is done within the stabilized path. The length stabilization therefore needs to operate in a pulsed way to remove shifts due to, both, slow long-term drifts and synchronous path length changes, e.g., due to mirror vibrations or switching induced phase chirps.

\begin{figure}
\begin{center}
\includegraphics[width=\columnwidth]{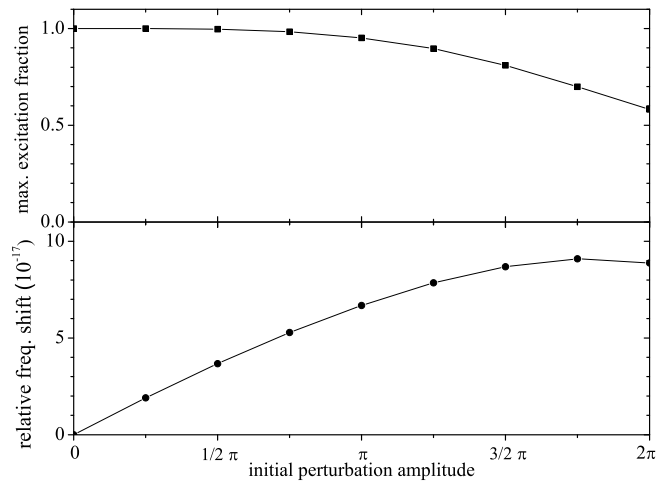}
\end{center}
\caption{\label{fig:vibrations}Simulated excitation probability and line shift of atoms interrogated with a phase modulated laser field (duration 90~ms). The phase~$\phi$ oscillates at 1~kHz damped with a time constant of 45~ms. The dependence of the maximal excitation fraction (top graph, $p_{\rm max}$) and the line shift (bottom graph) on the initial amplitude of the phase oscillations for our experimental conditions are shown.}
\end{figure}

\section{Perturbation}
Depending on the possible perturbation, the introduced frequency shift may average to zero over the time. If changes happen with a periodicity longer than an experimental run (typically a couple of hours) the frequency shift will not average out and may cause a net shift. Also, if an effect is synchronous with the interrogation pulse, the same phase chirp may be picked up every single interrogation and lead to a shift. A typical example for this are vibrations caused by laser beam shutters that lead to transient movement of mirror mounts. In order to estimate the possible effect of such vibrations, we calculated the atomic response using density matrix calculations discussed below. For an oscillation frequency of 1~kHz and damping times between 10~ms and 500~ms the resulting relative frequency shift can reach $10^{-16}$ even for oscillations as small as half a wavelength, where amplitude and sign of the induced shift depend on the timing of the vibrations with respect to the clock pulse. Only for larger amplitudes of the oscillations, the maximum excitation probability on the clock transition is reduced (Fig.~\ref{fig:vibrations}). Thus, a high excitation probability is not a sufficient indicator for having shifts caused by synchronous vibrations under control on the $10^{-17}$ level.

\section{Setup}
Details of the apparatus for cooling and trapping the reference atoms can be found elsewhere \cite{lis09,mid11,fal11}. Details of the clock operation are reviewed in Appendix~\ref{sec:clockReq}. The clock operation requires the use of different power levels and frequencies during each clock cycle, which had been considered in this work.

The clock laser providing the light to drive the clock transition is set up in a different room than the apparatus for the preparation of ultracold $^{87}$Sr atoms, which allows for a quieter and more temperature-stable environment. We use an extended cavity diode laser in Littman configuration and lock it to an ultra-stable high finesse reference cavity, leading to a linewidth of about 1~Hz \cite{vog11}. A bare laser diode is seeded with this light to amplify the optical power of the ultra-stable light. Its power is divided into two beams that are sent with polarization maintaining optical fibers to two sites: One beam is sent to the atoms, the other beam to a frequency comb, which allows to compare the frequency of the laser light to a primary cesium fountain clock at our institute. Moreover, frequency ratios of optical clocks can be measured with the comb directly. Length variations in the optical link to the comb are compensated by a usual fiber length stabilization \cite{ma94}.

\begin{figure}
\begin{center}
\includegraphics[width=\columnwidth]{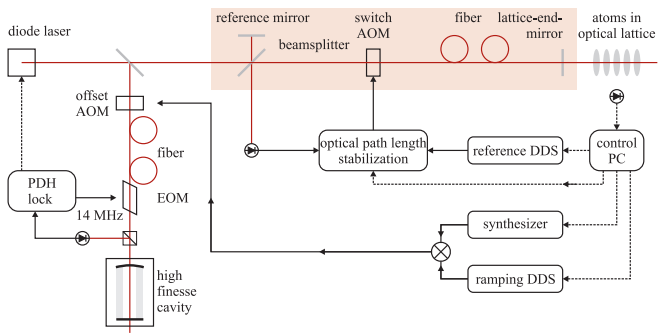}
\end{center}
\caption{\label{fig:lock}Three locks for the clock laser: A fast lock (Pound-Drever-Hall, PDH \cite{dre83}) to a cavity mode narrows the linewidth of the laser, a PC-controlled lock to the clock atoms that keeps the laser frequency long-term stable, and the optical path length lock discussed in this paper ensures that no systematic frequency shifts are introduced in the optical path between clock laser and interrogated atoms. The Michelson interferometer for path length stabilization is indicated by the shaded rectangle. Its reference mirror is used for a continuous fiber length stabilization (not shown) for the link to the frequency comb that counts the frequency of the clock laser, too.}
\end{figure}

Figure~\ref{fig:lock} shows the optical link between clock laser and reference atoms and the instruments for the optical path length stabilization. The light for probing the atoms is frequency shifted by the switch AOM and coupled into a 20 meter long polarization-maintaining fiber that connects to the neighboring laboratory. This AOM serves three purposes: It pulses the light for interrogating the atoms at controlled intensities, adds a variable offset for probing the line (App.~\ref{sec:clockReq}), and corrects for optical path length fluctuations, as explained below in detail. The length stabilization servos the phase of the rf signal driving the AOM to ensure a constant phase difference between light coming back through the fiber and a local reference. The local reference light is split off from the beam going into the fiber with a 10 percent beamsplitter and retro-reflected by the reference mirror, which serves as the end point of the short arm of a Michelson interferometer with unbalanced arm length. The other interferometer arm includes the AOM and the fiber (see shaded rectangle in Fig.~\ref{fig:lock}). In this arm, the clock laser light is retro-reflected by a mirror behind the fiber and re-traces the optical path: back through fiber and switching AOM back to the beam splitter of the interferometer. 

The optical path length stabilization ensures a constant optical phase difference between the light at the two end mirrors of the interferometer arms. Similarly, the cw link to the frequency comb fixes the phase at the comb to a local reference mirror. Thus, only the optical path between the two reference mirrors can cause instabilities or shifts at the two remote ends. Keeping this distance between the reference mirrors short or, as in our case, using the same mirror reduces its influence largely.

\begin{figure}
\begin{center}
\includegraphics[width=0.7\columnwidth]{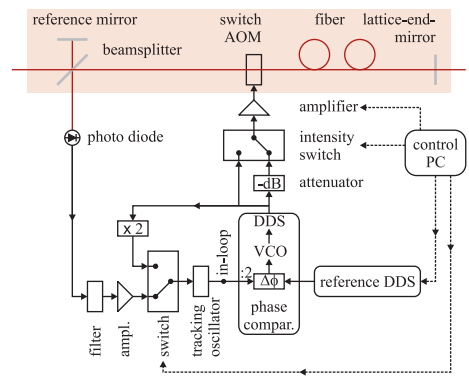}
\end{center}
\caption{\label{fig:stabilization}Details of the electronics of the optical path length stabilization between lattice-end-mirror and reference mirror. The beat note is detected with a photo diode, filtered, amplified, and tracked before being compared to the rf signal of the reference DDS. The switch between amplifier and tracking oscillator allows for a reduction of the frequency excursion of the VCO in the time when no beat note is present by using its frequency doubled rf signal instead.}
\end{figure}

The interference of the light retracing the path with fiber and AOM with light from the short local path creates a beat note at twice the AOM frequency (here around 160~MHz) since the frequency is shifted by the AOM on both ways in the same direction. The beat is detected with a fast photo diode with amplifier. A rf tracking oscillator is locked to the beat note and makes this signal available at higher signal-to-noise ratio for further analysis. As reference on the vacuum tank side of the experiment, we use light reflected from the same mirror coating that also retro-reflects the 813~nm optical lattice light on the other side. The standing wave pattern forming the optical lattice is automatically fixed to this lattice-end-mirror. Thus, (neglecting dispersion between 698~nm and 813~nm) the optical path length between reference mirror on the clock laser board and the atoms themselves is kept constant by the optical path length stabilization.

For the operation of the lattice clock the switching AOM needs to provide two power levels: high power for quick excitation $\pi$-pulses (4.5~ms, preparation) of the $^{87}$Sr atoms into a selected $m_{F}$ state of the $^3$P$_0$ level and low power for the actual clock pulse (90~ms, $\pi$-pulse if on resonance). This requires the optical power at the position of the atoms to be altered by a factor of 400 (26~dB), which is achieved by attenuating the rf power sent to the AOM. It is worth noting that the beat note of the path length stabilization is also attenuated by 26~dB at low intensity if compared to the beat note at the high intensity usually obtained at maximum AOM diffraction. In order to use high optical power and high AOM diffraction efficiency for the optical path length stabilization we chose a lattice-end-mirror that transmits just the right amount of clock laser light (0.1~\%) and reflects the rest, which is then used for the fiber length stabilization. Nevertheless, the optical power of the light arriving at the beat detector during the clock pulse is only in the order of a few nW while local reference light has a power of about 1~mW.

\begin{figure}
\begin{center}
\includegraphics[width=\columnwidth]{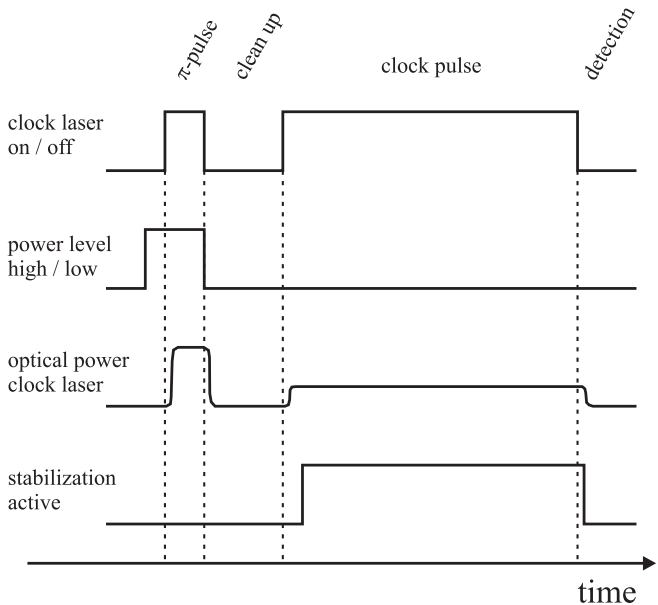}
\end{center}
\caption{\label{fig:timeScheme}Characteristic time scheme of the control of the clock laser pulses via TTL signals; not to scale. A short, high-power preparation pulse is followed by a short dark period and a longer, low power pulse for the clock interrogation. This pattern is repeated with a clock cycle time of 0.6~s.}
\end{figure}

As in cw path length stabilizations, any change in the length of the optical path between the two end mirrors of the Michelson interferometer translates into a phase shift between the frequency doubled rf signal driving the AOM and the beat signal. As shown in Fig.~\ref{fig:stabilization}, a digital phase comparator detects such changes by comparing the filtered, amplified, and tracked beat note, which is frequency divided by two, with a fixed rf frequency set to the nominal frequency offset of the light field between the two reference mirrors of the interferometer. The phase comparator output is fed to a voltage controlled oscillator (VCO) that clocks the direct digital synthesis (DDS) frequency generator (see Appendix~\ref{sec:extPhaseLock}). We ensured that the tracking oscillator used to increase the signal-to-noise ratio (see Fig.~\ref{fig:stabilization}) is not loosing cycles by splitting the amplified beat note signal and tracking it with two independent tracking oscillators. These two frequencies were divided by digital dividers. The open loop frequency of one VCO was below while the other VCO's frequency was above the beat note frequency in order to ensure that cycle slips cause a different behavior in both tracking oscillators that is observed in a change of the phase between the two dived signals. We verified that the phase of these signals is constant; therefore we have ensured that no cycle slips occur at the low intensity that we are operating with. In contrast to cw path length stabilizations, during the time between clock pulses no light and thus no signal is available for the optical path length stabilization. To avoid that the VCO and tracking oscillator frequencies are driven away due to the lack of the beat note, which would lead to long settling times at re-lock, we synthesize a beat signal by switching the input of the phase detector to the doubled frequency of the DDS output (see Fig.~\ref{fig:stabilization}).

The AOM introduces a delay between the applied rf voltage and light actually being diffracted of less than 2~$\mu$s (acoustic wave propagation in the AOM crystal, see Fig.~\ref{fig:timeScheme}). We switch on the optical path length stabilization after 16~$\mu$s. In other experiments, phase excursion within the first and last few $\mu$s of light pulses produced by AOMs have been observed \cite{deg05} but those are short enough to have only an effect in the $\mu$Hz regime for $\pi$-pulses of several 10~ms.

Small amplitude modulations occur only at the beginning of clock laser pulses and due to the small light shift associated with the clock laser field of $2 \cdot 10^{-17}$ \cite{fal11} are of no concern at the level of $10^{-18}$.

\section{Experimental Results}
Two aspects of the optical path length stabilization are of importance: Is the phase stabilized quickly enough after the switching of the clock laser? And: Is the lock leaving or---worse---introducing any systematic phase chirp across the clock laser pulse? In order to look into these issues, we have analyzed the in-loop signal (see Fig.~\ref{fig:stabilization}). Moreover, we generate an out-of-loop signal by looping the fiber back to the clock laser board and using a retro-reflecting optic on that board, where a beat with a local light beam is created. The filtering, amplification, and tracking of the beat signal is done in a similar way as for the in-loop signal. This beat note is at the AOM frequency since the light is diffracted only once by the AOM.

\subsection{Settling of Path Length Stabilization}

\begin{figure}
\begin{center}
\includegraphics[width=\columnwidth]{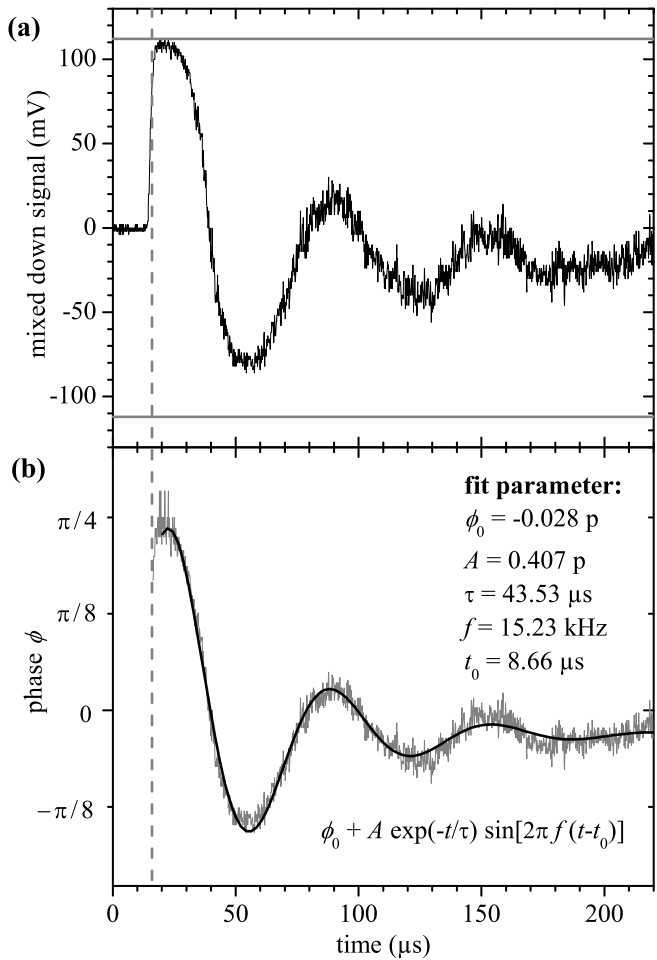}
\end{center}
\caption{\label{fig:start}Settling behavior of the optical path length stabilization. A delay is introduced between the TTL signal for switching the light on at 0~$\mu$s and the activation of the length stabilization: The dashed grey vertical lines indicate the activation time of the stabilization at 16~$\mu$s. In part~(a) voltage generated by mixing down the in-loop beat to the dc is shown for the switching-on of the active length stabilization. The grey horizontal lines indicate the signal for a $2\phi=\pi/2$ phase shift of the in-loop signal. In part~(b) the corresponding calculated phase excursion at the fundamental frequency $\phi$ is shown along with a fit as described in the text.}
\end{figure}

Both, the in-loop signal as well as the out-of-loop signal, show the same characteristics at the start of the stabilization: a few strongly damped oscillations (damping time 44~$\mu$s) of the phase that lead to a stable phase after about $200~\mu$s (see Fig.~\ref{fig:start}). Further optimization of the PI parameter did not shorten the damping time. This is possibly due to the over-all delay of the feedback loop.

From the observed single path residual phase excursion $\phi\left(t\right)$ one may calculate the corresponding frequency behavior
\begin{equation}
\delta\nu\left(t\right) = \frac{1}{2\pi} \frac{\rm d}{{\rm d}t}\phi\left(t\right)
\label{eq:freqFromPhase}
\end{equation}
during the clock pulse by taking the time derivative of $\phi\left(t\right)$. We use a parameterized damped oscillation function to describe the observed phase evolution:
\begin{equation}
\phi\left(t\right)=\phi_0+A\cdot \exp \left(-t/\tau\right) \cdot \sin  \left[2\pi f\left(t-t_0\right)\right]
\label{eq:phaseFit}
\end{equation}
where $\phi_0$ is the steady state phase, $A$ the amplitude at $t=0$, $f$ the servo oscillation frequency, and $\tau$ the damping time of the servo. We observed that the characteristic frequency and damping of the phase excursion behavior does not significantly depend on the phase at the time at which the lock starts operating. Over the course of several minutes the amplitude changes but $\phi$ remains within $\pm\pi/2$.

These phase excursion lead to changes of the excitation probability~$p$ leading to a shift of the locked laser frequency. Assuming a Lorentzian line of full width at half maximum FWHM and peak excitation probability~$p_{\rm max}$ a difference~$\Delta p$ of the two excitation probabilities at the half maximum points leads to a resulting frequency shift
\begin{equation}
\Delta \nu = \frac{\Delta p}{2\cdot p_{\rm max}}\ {\rm FWHM}\ .
\label{eq:shift}
\end{equation}

In the case of a Rabi excitation with a pulse duration~$T$ the slope at the half maximum points \cite{dic87} is 
\begin{eqnarray}
\frac{{\rm d}p}{{\rm d}\nu} & = & \pi\ 0.60386\ p_{\rm max}\ T = \pi\ 0.60386\cdot 0.798685\ \frac{p_{\rm max}}{\rm FWHM} \nonumber\\
& = & 1.516 \frac{p_{\rm max}}{\rm FWHM}\ ,
\end{eqnarray}
leading to a shift of
\begin{equation}
\Delta \nu = \frac{1}{1.516}\cdot\frac{\Delta p}{2\cdot p_{\rm max}}\ {\rm FWHM}
\end{equation}
and therefore Eq.~(\ref{eq:shift}) over-estimates frequency shifts. However, we use Eq.~(\ref{eq:shift}) for frequency shift estimations to account for possible additional errors, e.g., if one is not working exactly at the half maximum points but still in the steep region of the peak or one has other experimental imperfections that reduce the slope of the line otherwise.

We use two methods to obtain the change in excitation probability on one side of the line~$\delta p$ from the measured phase-excursion: first, we derive a sensitivity function and, second, we use a numerical calculation.

\subsubsection{Sensitivity Function}

\begin{figure}
\begin{center}
\includegraphics[width=\columnwidth]{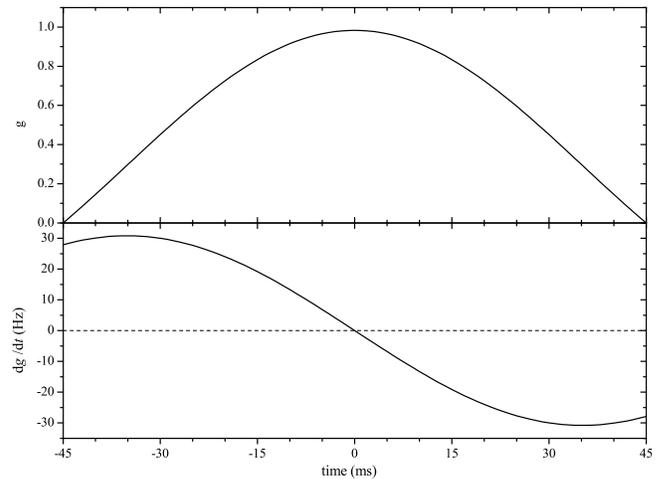}
\end{center}
\caption{\label{fig:sensitivity}The frequency sensitivity function $g$ (top) and its derivative (bottom) with respect to time ${\rm d}g/{\rm d}t$, which is the phase sensitivity function. The plots shown are calculated for a clock pulse of $T=90$~ms starting at $-45$~ms with a detuning~$\Delta$ of the half maximum point. This shows that the Rabi interrogation scheme is most sensitive to phase excursions at the beginning and the end of the clock pulse.}
\end{figure}

In general, the impact of a phase excursion on the excitation probability~$p$ depends on which part of clock laser pulse it is happening in. For a small frequency change~$\delta \nu(t)$ at time~$t$ the excitation probability changes by \cite{dic87,dic90,deg04}
\begin{eqnarray}
\delta p & = & \frac{1}{2}\int_0^T 2\pi\ \delta\nu\left(t\right) \cdot g\left(t\right) {\rm d}t \label{eq:sensFreq}\\
         & = & - \frac{1}{2}\int_0^T \phi\left(t\right) \cdot \frac{\rm d}{{\rm d}t} g\left(t\right) {\rm d}t, \label{eq:sensPhase}
\end{eqnarray}
where $g$ is the sensitivity function. According to Eq.~(\ref{eq:sensPhase}) its derivative can be interpreted as phase sensitivity function. In case of a $\pi$-pulse of length~$T$ centered at $t=0$ (equation 11 in \cite{dic87}) it is symmetric over the elapsed time~t:
\begin{eqnarray}
g\left(t\right) & = & \sin^2\vartheta \ \cos \vartheta \\
& & \times\left[\left(1-\cos\Omega_2\right)\sin\Omega_1\ + \ \left(1-\cos\Omega_1\right)\sin\Omega_2\right] \nonumber\\
{\rm with} & \vartheta & = \frac{\pi}{2}-\arctan \left( 2T \Delta \right) \nonumber \\
{\rm and} & \Omega_1 & = \pi \sqrt{1+\left(2T \Delta\right)^2}\times \frac{t}{T} \nonumber \\
& \Omega_2 & = \pi \sqrt{1+\left(2T \Delta\right)^2} \times\frac{T-t}{T} \nonumber
\end{eqnarray}
with $\Delta$ being the detuning of the clock laser from resonance (see Fig.~\ref{fig:sensitivity}). With a Rabi excitation by a clock pulse of length~$T$ the half maximum points are at $\Delta=\pm 0.399343/T$ according to literature \cite{dic87}.

We use the phase excursion shown in Fig.~\ref{fig:start}, which is parameterized by applying a fit according to Eq.~(\ref{eq:phaseFit}), to determine how our control deals with an initial phase difference between local oscillator and observed beat note. For a worst case scenario we assume that the phase initially is at $\phi=\pi/2$ from the start of the pulse until the activation of the control at 16~$\mu$s and then follows a damped oscillation with the observed characteristics. We derive the excursion of the frequency according to Eq.~(\ref{eq:freqFromPhase}) and do the integration of Eq.~(\ref{eq:sensFreq}). We obtain $\delta p = -4.2 \cdot 10^{-4}$. To estimate the frequency shift due to the residual phase chirp we assume that this change in excitation probability happens on both sides of the line with opposite sign, i.e. $\Delta p = 2\;\delta p\cdot p_{\rm max}$. With this, the frequency shift according to Eq.~(\ref{eq:shift}) is 3.8~mHz (FWHM~=~9~Hz, $p_{\rm max}=0.86$) for our clock laser running at about 429~THz.

The derivative of the sensitivity function with respect to time~$t$ is
\begin{eqnarray}
\frac{\rm d}{{\rm d}t} g\left(t\right) &= & \frac{2\pi \Delta}{1+\left(2 T \Delta\right)^2} \\
&&\times \left[\left(1-\cos\Omega_2\right) \cos\Omega_1 -\left(1-\cos\Omega_1\right)\cos \Omega_2\right] \nonumber 
\end{eqnarray}
Equal results were obtained by performing the integration according to Eq.~(\ref{eq:sensPhase}) directly using the observed phase excursions.

\subsubsection{Numerical Simulation}
\label{sec:simulation}
The use of a sensitivity function requires that the phase excursions can be treated as small perturbations. To avoid this limitation we take another approach to determine how the phase excursion alters the excitation probability. The excitation dynamics is modeled in a density matrix approach of the optical Bloch equations for a two-level atom \cite{fan57,coh92}.

If the two clock states and the coherence are represented by the Hermitian matrix $\rho$ (density matrix), their time dependence can be calculated according to
\begin{equation}
\frac{\rm d}{{\rm d}t}\rho=-\frac{i}{h}\left[H,\rho\right]-D
\label{eq:densMat}
\end{equation}
with the Hamiltonian
\begin{equation}
H=2\pi\left(
\begin{array}{cc}
0 & \Omega/2\cdot\exp\left(i\phi\right) \\
 \Omega/2\cdot\exp\left(-i\phi\right) & \Delta
\end{array}
\right)
\end{equation}
where $\Delta$ is the detuning of the clock laser, $\Omega$ the Rabi frequency at which the clock transition is driven at, and $\phi\left(t\right)$ is the instantaneous phase of the laser field. Additional dephasing due to the linewidth $\eta$ from white frequency noise of the laser is modeled by \cite{wod79}
\begin{equation}
D=2\pi \eta\left( 
\begin{array}{cc}
0 & \rho_{12}\\
\rho_{21} & 0\\
\end{array}
\right)\ ,
\end{equation}
which leads to a reduction of the coherence between the two levels.

Spectra are obtained by integrating Eq.~(\ref{eq:densMat}) starting with the population all in one state ($\rho_{11}=1$ and $\rho_{22}=0$ for $t=-\tau/2$) for different detunings~$\Delta$ and recording the final population in the second state, i.e., the transition probability~$p=\rho_{22}$ for $t=\tau/2$. For a perfectly stable link the phase excursion is $\phi =$~const at all times. We calculate the change of the excitation probability as in the worst case scenario discussed above. We obtain $\Delta p=1.1 \cdot 10^{-4}$, which corresponds to a shift of 0.6~mHz [Eq.~(\ref{eq:shift})]. This is significantly less than the result obtained with the sensitivity function. For very small phase excursions (less than $\pi$/10) the two approaches lead to similar results. If the phase excursion becomes bigger, the sensitivity function approach scales linearly while one observes a less-than-linear behavior and even sign changes with the density matrix approach. This difference can be explained by a violation of the assumption of small phase excursions that is essential for the sensitivity function approach.

The phase difference between the optical beat note of the link and the electronic shortcut that determines the initial phase and thus the phase amplitude may remain similar over many interrogation cycles. Therefore, as worst case for a frequency shift, we assume that this error does not averages out. But with $1\cdot 10^{-17}$, we have a good upper bound for the uncertainty introduced by the start of the length stabilization.

On the other hand, to estimate the influence of the initial phase fluctuations on the stability we assume as worst case scenario that this phase is uncorrelated between the interrogations. Then, our calculation of the frequency shift corresponds to the calculation of the Dick-effect in the time domain. Thus, this allows to estimate the influence of these fluctuations on the stability to be $\sigma_y(t_{\rm c}) < 1\cdot10^{-17}$ with the cycle time~$t_{\rm c}$.

The settling behavior of the servo loop can be further reduced by locking to the phase of the link not to zero but to its intial phase, i.e., the phase at the start of the current clock pulse.

\begin{figure}
\begin{center}
\includegraphics[width=\columnwidth]{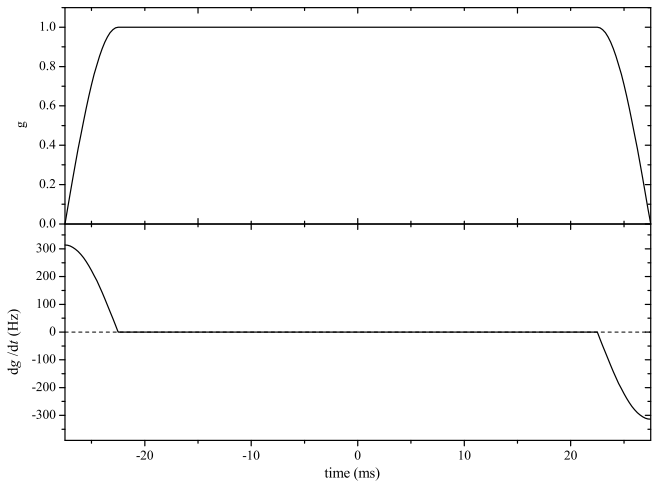}
\end{center}
\caption{\label{fig:sensitivityRamsey}The frequency and phase sensitivity function $g$ and ${\rm d}g/{\rm d}t$ for a Ramsey interrogation scheme with $\pi/2$-pulse of $t_{\rm i}=5$~ms and a dark time of $T_{\rm d}=45$~ms. This interrogation scheme is most sensitive to phase excursions at the beginning of the first pulse and the end of the second pulse. Note the change of scale compared to Fig.~\ref{fig:sensitivity}.}
\end{figure}

\subsection{Ramsey interrogation}
An alternative to the so far discussed Rabi excitation is a Ramsey interrogation scheme using several short more intense light pulses.

If one probes the transition in a Ramsey scheme with two $\pi/2$-pulses of interaction time $t_{\rm i}$ separated by a dark time of $T_{\rm d}=T/2$ (to obtain a similar FWHM as in the Rabi case) the frequency sensitivity function is \cite{dic87}
\begin{equation}
g\left(t\right)=\left\{
\begin{array}{ll}
\sin\left(\pi/2 \cdot t / t_{\rm i} \right) & {\rm for}\ 0 \le t < t_{\rm i}\\
1 & {\rm for}\ t_{\rm i} \le t < t_{\rm i} + T_{\rm d}\\
\sin\left[\pi/2 \cdot \left(t-t_{\rm i}-T_{\rm d} \right)/ t_{\rm i} \right] & {\rm for}\ t_{\rm i} + T_{\rm d} \le t < T_{\rm d} + 2t_{\rm i}\ ,\\
\end{array}
\right.
\end{equation}
as illustrated in Fig.~\ref{fig:sensitivityRamsey}. Both light pulses of the interrogation scheme will be accompanied by phase excursions similar to those shown in Fig.~\ref{fig:start}. The excursion of the first pulse is more important since the phase sensitivity function is small at the beginning of the second pulse. At the start of the interrogation scheme, the phase sensitivity function is approximately $\pi/(2t_{\rm i})$. In comparison, the phase sensitivity function for a Rabi excitation can be approximated by ${\rm d}g / {\rm d}t\approx 2.5 / T$ for half maximum detuning. In both interrogation schemes the sensitivity function changes much slower than the observed phase change during the activation of the length stabilization. Thus, as $t_{\rm i} \ll T$, a Ramsey scheme is more susceptible for phase chirps than a Rabi interrogation for the similar linewidths. However, one can slightly modify the pulses of the Ramsey scheme: Starting the (first) pulse with low light power, establishing the optical path length stabilization, and then increase to higher power while keeping the phase locked. For example, if one starts with the optical power of the Rabi excitation ($\pi$-pulse in $T=90$~ms, thus $t_{\rm i}=45$~ms) the change in the excitation probability is proportional to $\pi/T$, which is close to that of a Rabi excitation with $2.5/T$. An alternative approach to reduce the effect of residual phase excursions is discussion is Appendix~\ref{sec:hyper}.

\subsection{Slow Chirps}
In the previous section, we have assumed that the phase excursion settles to zero for longer times. There are certainly no big excursions at later times but we have to look at the possibility of residual slow drifts. An integrating part of the servo loop ensures that the servo error is pulled to zero. If it is too slow, e.g., if the time constant is longer or on the order of the length of the clock pulse, a long tail of the phase excursion over the clock pulse would remain. This would be seen as a frequency shift between the laser field seen by the atoms and the laser field counted by the frequency comb. This frequency shift might not average out in a measurement if, e.g., the phase lock starts pre-dominantly on one side of the lock point, which is likely the case for stable experimental conditions.

\begin{figure}
\begin{center}
\includegraphics[width=\columnwidth]{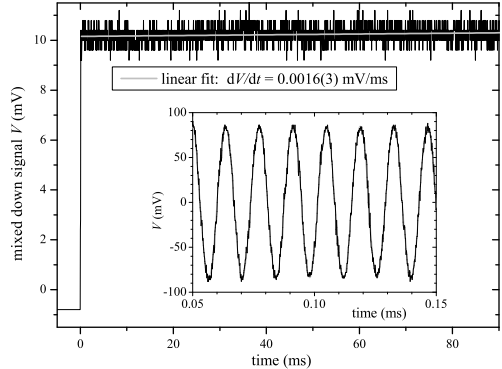}
\end{center}
\caption{\label{fig:drift}Residual phase drifts over the clock pulse. A phase of $\pi/2$ corresponds to a signal of $A=84$~mV as shown in the inset. The data shown is representing the average of 128 traces; the grey line shows a linear fit to the data.}
\end{figure}

In order to resolve even small drifts of the phase excursion we examine the out-of-loop signal. There is no lock of the optical path length in between the clock pulses, which leads to the problem that the lock using the double pass signal may lock to an even or odd multiple of $2\pi$ round trip phase $2\phi$ leading to 0 or $\pi$ of the single path signal, i.e., out-of-loop. We therefore frequency double the out-of-loop signal and mix it down with twice the reference frequency. Drifts in the uncompensated part of the optical paths are so significant that we need to adjust the phase of the reference oscillator manually in order to ensure that we stay in the steep region for the phase determination. In this way we avoid averaging of traces with opposing signs in the phase sensitivity. In Fig.~\ref{fig:drift} we show an example of an obtained curve and a linear fit to the data.

We measure the maximum amplitude of the mixed down beat signal~$A$ by switching off the optical path length stabilization (see inset of Fig.~\ref{fig:drift}). With this, we can relate the voltage~$V$ to the phase by looking at the zero crossing of the mixed down beat signal. The signal $V$ is periodic in the optical path length: assuming a change in the optical path length corresponding to a phase of $\phi$ (single way), the signals go as
\begin{equation}
V=A \sin\left(2\phi\right),
\end{equation}
where we have a factor of two due to the frequency doubling. From this we get that for the steepest part the phase change sensitivity is
\begin{equation}
\max \frac{{\rm d} V}{{\rm d}\phi} = 2A\ .
\label{eq:dVdphi}
\end{equation}
Not all of the averaged traces are close to the maximum slope. Therefore, we use a lower limit for the sensitivity ${\rm d}\phi / {\rm d}V <1.5/(2A)$ based on our experience with the range of observed phases in the manual control of the local oscillator phase. A maximal residual frequency offset~$\Delta\nu$ due to the phase chirp across the full clock pulse---measured by the slope ${\rm d}V / {\rm d}t=1.5$~mV/s---can then be calculated to be
\begin{equation} 
\Delta\nu=\frac{{\rm d}\phi}{{\rm d} t}\cdot\frac{1}{2\pi}=
{\frac{{\rm d}\phi}{{\rm d}V} \cdot \frac{{\rm d}V}{{\rm d} t}\cdot\frac{1}{2\pi}}
\end{equation}
For the trace shown in Fig.~\ref{fig:drift}, the upper limit of the resulting frequency shift due to residual drift of the phase would be 2.4~mHz. This corresponds to an uncertainty of a frequency measurement due to a residual linear drift of the optical path length of $6\cdot 10^{-18}$.

\section{Summary}
We have set up an optical path length stabilization for clock laser pulses of a strontium lattice clock at a frequency of 429~THz. The phase of the laser field at the position of a retro-reflecting mirror is stabilized within 200~$\mu$s to $\pm40$~mrad. This mirror is the retro-reflecting mirror of an optical lattice, in which the atoms of the optical clock are trapped. In this way, the optical path between clock laser board and reference atoms is stabilized. The stabilization uses an AOM within the stabilized path, which is also used for switching the clock laser. Preparation pulses with the clock laser at higher power than for the interrogation are possible, too. The fractional frequency shifts due to residual phase excursion at the start are below $10\cdot 10^{-18}$ and slow drifts are below $6\cdot 10^{-18}$. This path length stabilization has been proven to be reliable in a frequency measurement lasting several hours \cite{fal11}. The detailed analysis of our setup allows for giving a reliable upper bound for effects related to optical path length changes regardless of their source: vibrating mirrors and temperature related drift or chirps in AOMs are all compensated for at the level of $2\cdot 10^{-17}$. The fiber length stabilization also removes degradations of the clock stability to a level of $1\cdot 10^{-17}$ in a single shot.

In the future, laser with better short-term stability may allow for longer interrogation times~$T$. With the present optical path length stabilization, both, uncertainty and instability will scale down approximately as $1/T$. Our pulsed optical path length stabilization may be applied in any optical clock setup. Especially in optical \emph{lattice} clocks the optical path length stabilization links the clock laser directly to the position of the interrogated atoms leaving no uncompensated optical path.

\appendix
\section{Clock operation requirements}
\label{sec:clockReq}
The narrow linewidth of the clock laser light is obtained by a lock of the master laser to an ultra low expansion glass (ULE) reference cavity (see Fig.~\ref{fig:lock}). A double pass AOM allows for a frequency offset between laser light and cavity mode. A polarization maintaining fiber is used between this AOM and the cavity. The lock to a TEM$_{00}$ mode of the cavity is done with a Pound-Drever-Hall method, which uses a 14~MHz electro optical modulator (EOM) to create sidebands. The obtained servo signal is fed back to the current of the master diode and at low frequencies to the piezo of the grating for long-term corrections. 

As for most ULE cavities, the length of the reference cavity is shrinking slowly \cite{dub09}. In our case the frequency drift is around 30~mHz/s. To ensure that the laser keeps the same frequency while staying locked to the cavity we apply a ramp to the frequency offset AOM. In order to lock the laser to the atomic reference, we also apply frequency corrections to the AOM, which are determined by the control PC. We use a home built frequency generator for the drift (see Appendix~\ref{sec:ramp}) and a synthesizer for the corrections. The AOM is driven with the sum frequency of these two sources, which is obtained from a mixer as indicated in Fig.~\ref{fig:lock}.

To operate the optical clocks, i.e., to servo the laser to the line center of the reference transition, the line is probed at the half maximum points. Because of hyperfine structure in $^{87}$Sr the clock transition is split into several Zeeman components and the laser needs to be locked to the average of the two extreme Zeeman components \cite{fal11}. A Zeeman component and the side of the fringe is addressed through altering the frequency at the end of the fiber link by changing the reference frequency of the noise cancellation. This frequency shift of few kHz is applied by the switching AOM, thus the laser itself runs all the time at the same frequency, which is beneficial for the frequency counting or frequency comparison with the frequency comb. Before the actual clock interrogation, a clock laser pulse excites the atoms to a specific $m_F$ state of the $^3$P$_0$. During this preparation $\pi$-pulse the magnetic field is higher than during the clock laser interrogation in order to split the hyperfine transitions more, which allows spectrally resolving the hyperfine structure even with shorter pulse duration. The frequency driving the switching AOM is altered accordingly within a clock cycle to account for the different magnetic field. After the preparation pulse undesired atoms in the $^1$S$_0$ level are removed with resonant light and the magnetic field is reduced before the clock laser pulse starts.

\section{Frequency generation}
Throughout the experiments, various radio frequencies are generated and sent to AOMs. We have combined a DDS chip (Analog Devices AD9956 \cite{ana04}) with a micro-controller (Atmel ATmega644) and created a versatile rf source. This source may operate in different modes: firstly, creating a ramp in the frequency, secondly, phase coherent switching of frequency and phase, and thirdly, a mode that allows for using it as a VCO that can be controlled via a phase detector. These modes are discussed in more detail below. The micro-controller is controlled from a PC by a USB interface, which is provided by an on-board USB to RS-232 interface. A set of commands is supported to process requests.

The clock signal of the DDS chip is provided by a low phase noise 400~MHz VCO (Vectron VS-500). Its output signal can be phase locked to a reference with a phase detector and a frequency divider, both provided by the DDS chip. For the first two modes we lock the 400~MHz VCO to an external 100~MHz reference.

This phase locked loop (PLL) has a unity gain bandwidth of 15~kHz, above which the phase noise is determined by the phase noise of the VCO. For a carrier frequency of 100~MHz the phase noise at $f=15$~kHz is at $-100$~dBc and falls off quickly towards higher frequencies. Below that point, the system inherits the phase noise of the reference and adds inherent noise. In the range 10~Hz $<f<$ 15~kHz the inherent noise decreases from $-90$~dBc at 10~Hz approximately as $f^{-1}$. For even lower frequencies the inherent noise remains constant at $-90$~dBc.

\subsection{Frequency Ramp}
\label{sec:ramp}
The generation of a frequency ramp with a AD9956 is one if its on-chip features. But the resolution of the sweep rate is too coarse for a cavity drift compensation. We therefore generate the frequency ramp of our rf source by changing the output frequency of the DDS by the micro-controller at a rate of about 100~Hz while maintaining the phase coherence. The frequency resolution of the DDS is 48-bit, i.e., 1.4~$\mu$Hz, while the phase resolution is 14-bit. The sole input to the micro-controller is the start frequency and the rate of the frequency ramp. The micro-controller also allows for the read-out of the instantaneous frequency via the USB link. A frequency generator running in this mode is used to compensate for the linear drift of the reference cavity of the laser via an AOM.

\subsection{Frequency Switch}
The DDS chip keeps eight profiles with pairs of frequency and phase offset. One may select one of these profiles by three TTL signals. When switching between profiles the phase is altered by the difference between the two phase offsets and the frequency is changed. If the phase offsets are the same, the frequency switches while the phase of the output remains continuous. In this mode, the micro-controller is used to set the values of the frequencies and phase offsets via the USB port. The selection of the profile (frequency and phase) is done purely by the TTL signals, which allows for $\mu$s timing. In our experiment, this mode is used to quickly alter the clock lasers frequency, i.e., between preparation and clock pulse or---less time critical---for addressing the Zeeman components.

\subsection{External Phase Lock}
\label{sec:extPhaseLock}
This mode is used for the rf source driving the AOM of the optical path length stabilization as described in this work. It uses the phase detector and divider stages of the AD9956 to generate a signal that is fed to the tuning input of the 400~MHz VCO. In this mode the input to the phase detector are two external signals. In our case of the path length stabilization these are the beat note and the rf reference. The pull range of the VCO is $\pm 5 \cdot 10^{-5}$, which allows to detune the AOM frequency of 80~MHz by several kHz. 

\section{Hyper Ramsey}
\label{sec:hyper}
\begin{figure}
\begin{center}
\includegraphics[width=\columnwidth]{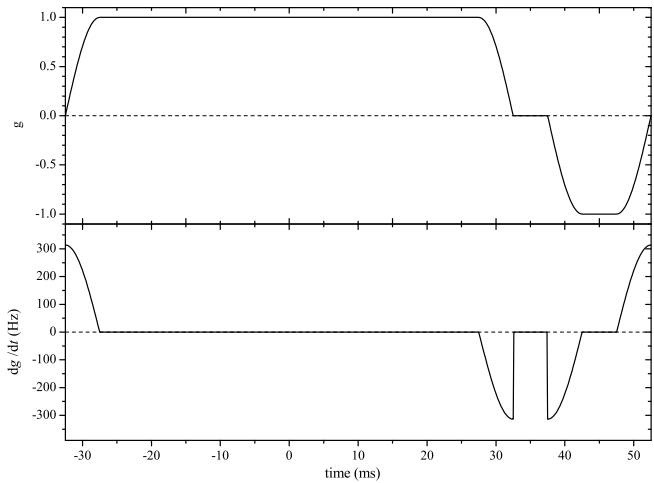}
\end{center}
\caption{\label{fig:hyper}Frequency and phase sensitivity for a hyper Ramsey like excitation scheme. The dark time in this scheme needs to be longer than in a Ramsey scheme to obtain the same sensitivity. This is due to the sign change in $g(t)$ between the second and the third pulse: the additional pulses reduce the sensitivity of a Ramsey scheme (first two pulses only) but allow for an automatic compensation of effects related to the activation of the optical path length stabilization.}
\end{figure}

A method to compensate shifts introduced by the interrogation pulses is the application of a hyper Ramsey scheme \cite{yud10}, where a $3\pi/2$ pulse is used as the second interaction pulse. Similar to this scheme we propose an interrogation scheme with one $\pi/2$ pulse in beginning and, after a long dark time, three individual $\pi/2$ pulses. Two short dark times between the three pulses reset the optical path length stabilization to have a similar phase chirp in all four pulses. This sequence has a phase sensitivity function (see Fig.~\ref{fig:hyper}) that may help to reduce the influence of the observed phase excursions and the investigated slow phase drifts. If one assumes that the phase behaves exactly the same for each of the four pulses their effects cancel as the integral of the product of phase sensitivity function and assumed phase excursion is zero. The effect from the first pulse is canceled by the third and the second is canceled by the fourth pulse.

\begin{acknowledgments}
The support by the Centre of Quantum Engineering and Space-Time Research (QUEST), funding from the European Community's ERA-NET-Plus Programme (Grant No. 217257), from the European Community's Seventh Framework Programme (Grant No. 263500), and by the ESA and DLR in the project Space Optical Clocks is gratefully acknowledged. We thank Burghard Lipphardt for helpful discussions. 
\end{acknowledgments}

\bibliographystyle{nar}

\end{document}